\documentclass{aa}  

\usepackage{graphicx}
\usepackage[varg]{txfonts}
\usepackage[normalem]{ulem}
\usepackage[usenames]{color}
\usepackage{mathrsfs}
\usepackage{natbib}
\usepackage{enumitem}

\usepackage{geometry}                		
\usepackage{graphicx}				
\usepackage{amssymb}
\usepackage{nccmath}
\usepackage{booktabs}
\usepackage[tableposition=top]{caption} 
\usepackage{multirow}
\usepackage{hyperref}
\usepackage[font=small,labelfont=bf]{caption}
\usepackage{media9}
\newcommand{\Mm}{{\mathrm{\, Mm}}}
\newcommand{\kms}{{\mathrm{\, km \,s^{-1}}}}

\newcommand{\secs}{{\mathrm{\, seconds}}}
\newcommand{\mins}{{\mathrm{\, minutes}}}
\DeclareMathOperator\erf{erf}

\begin{document} 
\title{Numerical simulations of large-amplitude oscillations in flux-rope solar prominences}

   \author{V. Liakh\inst{1, 2}
   \and 
   M. Luna \inst{3, 4}
   \and 
   E. Khomenko\inst{1, 2}}

 \institute{Instituto de Astrof\'{\i}sica de Canarias, E-38205 La Laguna, Tenerife, Spain\\
              \email{vliakh@iac.es}
         \and
             Departamento de Astrof\'{\i}sica, Universidad de La Laguna, E-38206 La Laguna, Tenerife, Spain\\
         \and
             Departament de F\'{\i}sica, Universitat de les Illes Balears, E-07122, Palma de Mallorca, Spain\\
         \and
             Institut d'Aplicacions Computacionals de Codi Comunitari (IAC$^3$), Universitat de les Illes Balears, E-07122, Palma de Mallorca, Spain
             }
   

 \date{}

 \titlerunning{Numerical simulations of LAOs}
\authorrunning{Liakh, Luna \& Khomenko}

\abstract
   {Large-amplitude oscillations (LAOs) of the solar prominences are very spectacular but poorly understood phenomena. These motions have amplitudes larger than  $10 \kms$ and can be triggered by the external perturbations, e.g., Moreton or EIT waves.}
   {Our aim is to analyze the properties of large-amplitude oscillations using realistic prominence models and the triggering mechanism by external disturbances.}
   {We perform time-dependent numerical simulations of LAOs using a magnetic flux rope model with two values of the shear angle and the density contrast. We study the internal modes of the prominence using the horizontal and vertical triggering. In addition, we use the perturbation that arrives from outside in order to understand how such external disturbance can produce LAOs.}
   {The period of longitudinal oscillations and its behavior with height show good agreement with the pendulum model. The period of transverse oscillations remains constant with height, suggesting a global normal mode. The transverse oscillations typically have shorter periods than the longitudinal oscillations.}
   {The periods of the longitudinal and transverse oscillations show only weak dependence on the shear angle of the magnetic structure and the prominence density contrast. The external disturbance perturbs the flux rope exciting oscillations of both polarizations. Their properties are a mixture of those excited by purely horizontal and vertical driving. 
   }

 
   \maketitle
%

 %

\section{Introduction}

Solar prominences (filaments when observed on the disk) are giant clouds of cold plasma suspended in the solar corona.
The prominences are very dynamical structures subject to complex motions, including different types of oscillations. These oscillations have been classified into small- and large-amplitude oscillations. Large-amplitude oscillations (LAOs) properties' differ much from the small-amplitude oscillations. LAOs have velocities larger than $10 \kms$ and often triggered by the external perturbations as Moreton \& EIT  waves \citep{Eto:2002pasj, Gilbert:2008apj, RiuLiu:2013apj} or  EUV waves \citep{Liu:2012apj, Shen:2014apj1, Zhang:2018apj}. In LAOs, a large portion of the filament oscillates, reflecting global characteristics of the plasma and the magnetic field structure. In contrast, small-amplitude oscillations are probably not excited externally, involving only a small part of the prominence \citep [see review by][]{Arregui:2018lr}. The LAOs can be additionally classified according to the polarization of the oscillatory motions. The longitudinal LAOs are motions parallel to the magnetic field, whereas the transverse LAOs are motions perpendicular to the magnetic field. In turn, the transverse oscillations can be vertically or horizontally polarized. Detailed information on the prominence oscillations, their classification, and properties can be found in reviews by \citet{Tripathi:2009ssr} and, more recently, by \citet{Arregui:2018lr}.

Since the first observations of the large-amplitude longitudinal oscillations (LALOs) by \citet{Jing:2003apjl, Jing:2006solphys}, the number of the reported events has increased \citep[e.g.][]{Zhang:2012aa, Li:2012apjl, Luna:2014apj}. LALOs are characterized by a period of about one hour; the direction of these motions coincides with the direction of the magnetic field, forming an angle of approximately $25^{\circ}$ with respect to the filament spine \citep{Luna:2014apj, Luna:2018apjs}. This value of angle is in agreement with direct magnetic field measurements \citep[e.g.][]{Leroy:1983solphys,Leroy:1984aa,Casini:2003apj}. Several candidates have been proposed as a restoring force for the LALOs. \citet{Luna:2012apj} proposed the so-called pendulum model where the main restoring force is gravity projected along the magnetic field, and the period depends only on the radius of curvature of the dipped magnetic field lines where the cool prominence plasma resides. \citet{Terradas:2013apj}, based on 2D magnetohydrodynamics (MHD) simulations, showed that the gas pressure gradient could make a significant contribution to the restoring force under some circumstances. \citet{Luna:2012apj1} have shown that the restoring force is a combination of the gravity projected along the field lines and the gas pressure gradient, but under realistic prominence conditions the gravity is the dominant force.
Numerical 2D and 3D MHD simulations \citep{Luna:2016apj, Zhou:2018apj} have confirmed that gravity is the main restoring force in agreement with the pendulum model. However, the above numerical experiments are limited to a just single prominence model. It is desirable to investigate how the oscillatory parameters depend on the prominence properties by considering several prominence models.

The transverse oscillations of both vertical \citep{Ramsey:1965aj, Ramsey:1966aj, Eto:2002pasj, Okamoto:2004apj, Gilbert:2008apj, Zhang:2018apj} and horizontal polarizations \citep{Isobe:2006aa, Hershaw:2011aa} have been also observed. These oscillations usually have a short period between $10\mins$ and $20\mins$. From a theoretical point of view, transverse oscillations were studied first in the works of \citet{Hyder:1966zap} and \citet{Kleczek:1968solphys}. The prominence was considered as a harmonic oscillator with the magnetic tension as the restoring force. Later, the properties of the transverse short-period oscillations have been studied using 2D and 3D numerical simulations \citep{Terradas:2013apj, Luna:2016apj, Zhou:2018apj}. \citet{Zhou:2018apj} have concluded that the magnetic tension is a dominant restoring force for both vertical and horizontal short-period oscillations.\citet{Terradas:2013apj} and \citet{Luna:2016apj} have found that the vertical period remains constant with height, indicating that it is a global normal mode. Unlike that, \citet{Terradas:2013apj} suggested that the vertical period has a simple dependence on the properties of the plasma and the magnetic field. The period increases with the prominence density and decreases with the length of the field line. For prominence seismology, the relation between the vertical period and the properties of the global magnetic structure can provide valuable information on the prominence structure.

As for the excitation mechanisms of LALOs, many observations reported LALO triggering by the disturbance close to the filament. In contrast, transverse LAOs are associated with the distant flares and their corresponding EIT/Moreton waves \citep[see review by][]{Arregui:2018lr}. However, observations have also shown that different polarizations can be excited simultaneously by external disturbances. 
\citet{Gilbert:2008apj} have reported the event of the simultaneous longitudinal and transverse LAOs triggered by a Moreton wave. Using the Atmospheric Imaging Assembly (AIA) data from the Solar Dynamics Observatory \citep{Lemen:2012solphys} and the SMART H$\alpha$ data \citep{Ueno:2004}, \citet{Asai:2012apjl} have found that the prominence oscillates in different directions simultaneously due to the external perturbation. \citet{Shen:2014apj} have reported the excitation of the transverse oscillations of one filament and the longitudinal oscillations of another filament triggered by the same shock wave. The authors suggested that the motion type strongly depends on the propagation direction of the shock wave with respect to the prominence axis. Despite all the efforts, it is still unclear how the energetic EIT/Moreton disturbances propagate from a flare and produce the oscillations of different polarizations in a distant filament.

In this work, we aim to understand the LAOs in a realistic prominence structure and how the LAOs properties' depend on different parameters of the structure (i.e., density contrast, shear angle). Additionally, we study how an external disturbance excites different modes of oscillation of the structure.

This paper is organized as follows: in Section \ref{Numerical}, an initial numerical configuration is described. In Section \ref{Internal}, it is explained how the prominence was perturbed and how the plasma and magnetic field respond to the internal (horizontal and vertical) and external perturbations. In Section \ref{Conclusions}, the main results are summarized.

\section{Numerical model}\label{Numerical}

Our objective is to study the oscillatory properties of the prominence models triggered by internal and external perturbations. We solve numerically the equations of ideal MHD using the state-of-the-art Mancha 3.0 code \citep{Khomenko:2008solphys,Khomenko:2014aa,Felipe:2010apj,Khomenko:2012apj}. We used the 2.5D approximation where all the vectors have three spatial components, but the perturbation is only allowed to propagate in two dimensions. The governing equations and the corresponding source terms are described by \citet{Felipe:2010apj}.
In this section, the numerical setup is outlined. We form the prominence structure by the cancelation process from a force-free arcade. Once the flux rope is formed, we load the cool prominence mass.
The initial configuration and the processes of the flux rope formation, mass loading and the relaxation are explained in Sections \ref{subsec:Initial equilibrium}, \ref{subsec:Formation} and \ref{subsec:Mass loading}, respectively.
The numerical domain consists of $800\times 600$ grid points, which corresponds to the physical size of $192\Mm \times\ 144 \Mm$ (a spatial resolution of $0.24\Mm$), except for one case where we used four times higher resolution. The Cartesian coordinate system is used with $z$- and $x$-axes defining the vertical and horizontal directions, respectively. 

\subsection{Initial equilibrium}\label{subsec:Initial equilibrium}

The initial configuration consists of an isothermal atmosphere with a temperature of $T=2\ $ MK gravitationally stratified along the vertical direction with a density at the bottom of the numerical domain of $\rho=1.31\times 10^{-12}\ \mathrm{kg\ m^{-3} }$. We have used an initial temperature larger than the typical value of the low corona of 1 MK. The reason for this selection is to have a large pressure scale height. This allows us to have smoother stratification, avoiding huge Alfv\'{e}n speeds at the top of the domain. The thermal conduction is not taken into account, but the radiative losses will be included during the stage of the flux rope formation later on.  We consider a monoatomic gas with an adiabatic constant, $\gamma=5/3$. Using these values of the thermodynamical parameters, we have obtained the maximum sound speed, $c_s=\sqrt{\gamma p_0/\rho_0}=166\kms$, and density scale height, $H=47 \Mm$, in the initial atmosphere.

The initial force-free magnetic field is a periodic sheared arcade defined as:
\begin{eqnarray}\label{magnetic_field_component_x}
B_{x}&=&\frac{2LB_{0}}{\pi H_{b}}\cos (k(x-x_{0}))\exp\left({-\frac{z-z_{0}}{H_{b}}}\right) \, , \\ \label{magnetic_field_component_y}
B_{y}&=&\sqrt{1-\left(\frac{2L}{\pi H_{b}}\right)^2}B_{0}\cos (k(x-x_{0}))\exp\left({-\frac{z-z_{0}}{H_{b}}}\right) \, ,\\ \label{magnetic_field_component_z}
B_{z}&=&B_{0}\sin (k(x-x_{0}))\exp\left({-\frac{z-z_{0}}{H_{b}}}\right) \, ,
 \end{eqnarray}
where $(x_0, z_0)=(0, 1 )\Mm$ are the coordinates of the center of the magnetic structure. The polarity inversion line (PIL), in this case, coincides with the line $x=0$.  The parameter $L$ is related to the lateral extension  $k=2\pi/L$, and the constant $A$  determines the magnetic scale height. In our experiments, we have considered $L=48\Mm$ and $B_0=20$ G.

\begin{figure*}[!ht]
\centering\includegraphics[width=0.9\textwidth]{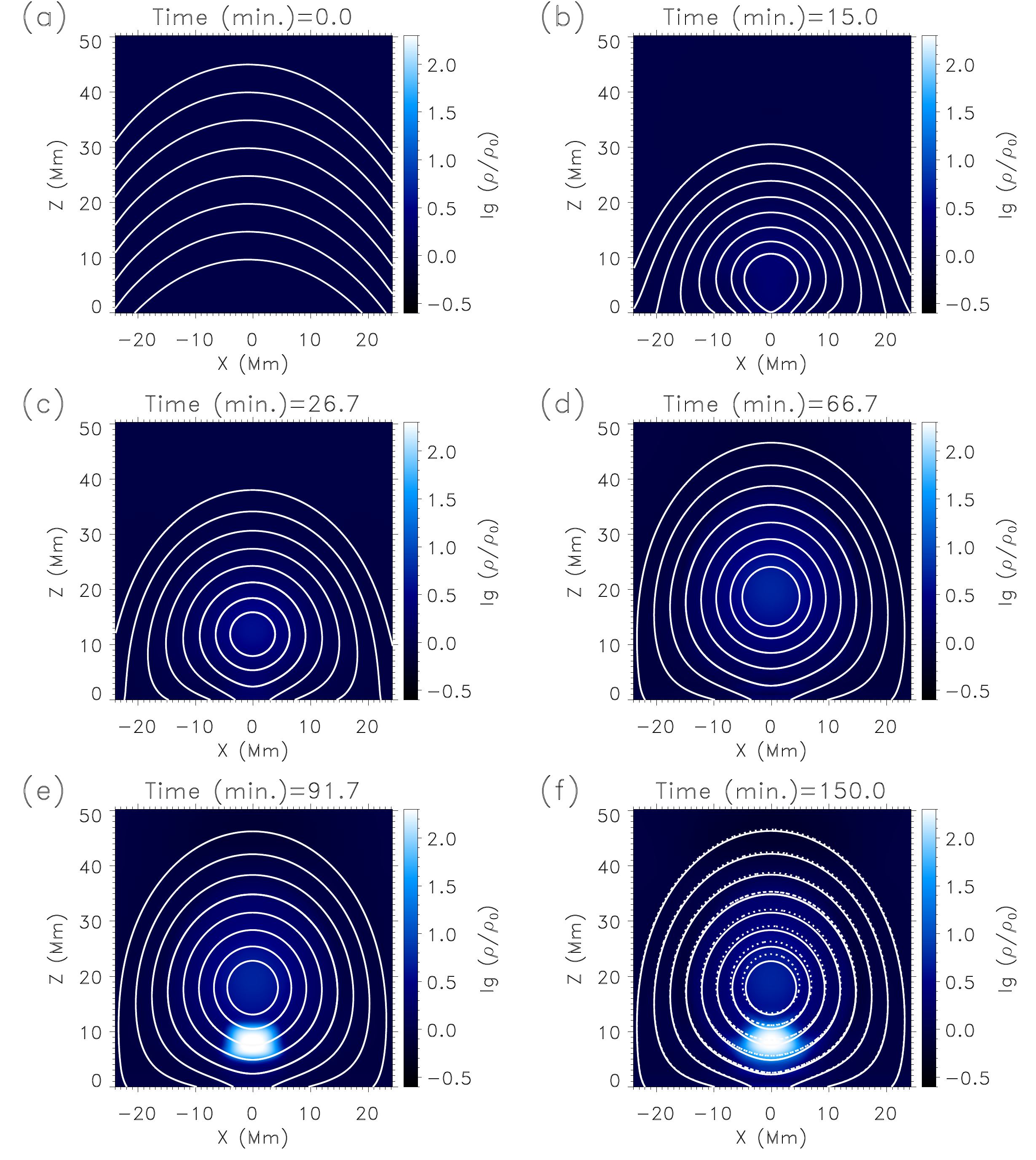}
\caption{The magnetic field lines and the density distribution in the central part of the computational domain during the prominence formation. Panel (a): the initial atmosphere and the magnetic field. Panels (b)-(d): the processes of the convergence and the reconnection. Panel (e), (f): the initial and final stages of the mass loading. At panel (f), the dashed line marks the magnetic field lines just before the mass loading process. The color bar scale indicates the logarithm of the density contrast.\label{fig:formation}}
\end{figure*}

It is useful to define an initial shear angle, $\theta$, as
 \begin{equation}\label{shear_angle}
     \tan\theta=\frac{B_{y}}{B_{x}}=\sqrt{\left(\frac{\pi H_{b}}{2L}\right)^2-1} \, ,
 \end{equation}
that is the angle between the arcade field and the line perpendicular to the PIL. A large angle $\theta$ involves large magnetic shear. From Equation \eqref{shear_angle}, we see that there is a relation between the shear, $\theta$, and the magnetic stratification, $H_{b}$. We consider the models with two initial shear angles, $\theta=60\,^{\circ},\ 45\,^{\circ}$. The corresponding parameters of the structure have values of $H_{b}=61,\ 43 \Mm$, respectively. The initial magnetic field in the 2D projection is shown in Figure \ref{fig:formation}(a). In all the initial configurations considered the maximum Alfv\'{e}n speed is $v_\mathrm{A}=B/\sqrt{\mu_0\rho_0}=1594\kms$ and the plasma-$\beta$ less than one everywhere in the domain.

\subsection{Flux rope formation}\label{subsec:Formation}
Observations show that cool and dense prominence plasma is usually located in the dipped part of the curved magnetic field (see, e.g., reviews by \citet{Mackay:2010ssr} and \citet{Vial:2015})
The dense prominence plasma is supported against gravity by the Lorentz magnetic force \citep{Kippenhahn:1957zap, Kuperus:1974aa}. Several models have been proposed for the magnetic structure of prominences, but the flux rope structure is the most plausible.
In this work, we consider the formation of the flux rope from the initial sheared arcade using the \citet{vanBallegooijen:1989apj} mechanism.
This approach is similar to \citet{Kaneko:2015apj}. We have only applied the converging motions at the bottom boundary in direction toward the PIL. The horizontal velocity imposed at the base is
\begin{equation}\label{imposed_velocity}
V_{x}(x, t)=-V_{0}(t)\sin{\left[\frac{2\pi(x-x_{0})}{L}\right]}\exp{\left[-\frac{(x-x_{0})^2}{2\sigma^2}\right]} \, ,
\end{equation}
where parameter $L=189.6 \Mm$ is the double width of the arcade, $x_0=0 \Mm$ is the position of PIL, $\sigma=13.6 \Mm$ is related to the half-size of the converging region. We additionally set the $V_y$ and $V_z$ velocity components to zero at the bottom of the domain. 
We activate and desactivate the cancelation process smoothly with a function $V_0(t)$ defined as
 \begin{eqnarray}\nonumber
 V_0(t)= V_{00}  \left\{ 0.5\left[\erf\left(\frac{t-2\lambda-t_1}{\lambda}\right)+1\right] \right. \\ \label{temporal_evolution}
   \left.-0.5\left[\erf\left(\frac{t-2\lambda-t_2}{\lambda}\right)+1\right] \right\}  \, ,
 \end{eqnarray}
where ${V_{00}=3V_{max}}$. The parameter $V_{max}=38\kms$ is the maximum converging velocity. Since we are interested in the final result of the flux rope formation, and not in the process, we used this large value of velocity to achieve the final stage quickly. The parameters $t_1=100\secs$ and $t_2=3100\secs$ determine the initiation and the termination of the formation process. The parameter $\lambda=150\secs$ determines the duration of the transition, and consequently, it is responsible for the smoothness of the cancelation process.
The cancelation process produces a heating of the plasma of the flux-rope. Consequently, the gas pressure inside of the rope and the plasma-$\beta$ also increase to values larger than desirable in a prominence \citep[see, e. g.,][]{Hillier:2012apj}. This effect was already described by \citet{Kaneko:2015apj}. Following these authors, we include a radiative loss term in our energy equation \cite[Eq. (3) from][]{Felipe:2010apj}. We use Newton's radiation law given by
\begin{equation}
    Q_\mathrm{rad}=-c_v \, \frac{T_1}{\tau_\mathrm{R}} \, ,
\end{equation}
where $T_1$ is the perturbation of the temperature with respect to the initial 2 MK, $\tau_\mathrm{R}$ is the radiative relaxation time, and $c_v$ is the specific heat at constant volume. This $Q_\mathrm{rad}$ is only used during the flux-rope formation process. We use a spatially constant, and very small, $\tau_\mathrm{R} = 3$ seconds, in order to keep the temperature near the $2$ MK. At the end of the cancelation, the temperature is $2$ MK, and then we deactivate the $Q_\mathrm{rad}$ term.
In Figure \ref{fig:formation}(a)-\ref{fig:formation}(d) different stages of the formation are shown. Figure \ref{fig:formation}(a) shows the initial field of the magnetic arcade, described by Equations \eqref{magnetic_field_component_x}-\eqref{magnetic_field_component_z}. Figure \ref{fig:formation}(b) shows how the footpoints of the field lines move toward the $X=0$ axis following Equation \eqref{imposed_velocity}. When the field lines approach each other, they reconnect near the origin of the coordinates, forming a twisted magnetic structure (Fig. \ref{fig:formation}(c)). The flux rope slowly evolves through a series of stable equilibria and rises to the higher heights. The continuation of the converging process at the bottom could eventually lead the system to lose equilibrium and erupt. However, in the present work, we do not study the stability and eruption of the prominences. This problem can be addressed in future work. To avoid the eruption and to have a stable prominence, we stop the convergence at $t=55\mins$. The resulting flux rope (see Fig. \ref{fig:formation}(d)) has the center at the height of $18.24\Mm$. The magnetic field strength in the dipped part is about of $14$ G. The characteristics of the flux rope are comparable to those observed in the solar prominences.

We use symmetric (for $V_{x}, V_{y}, B_{z}$) or anti-symmetric (for $V_{z}, B_{x}, B_{y}$) boundary conditions at the top boundary. The gas pressure and the density are fixed by assuming hydrostatic equilibrium at a constant temperature. The periodic boundary conditions are assumed at the left and the right boundaries. At the bottom, we let the magnetic field to evolve according to the velocity field. During the flux rope formation process, the magnetic field has been computed from the induction equation using the converging velocity. Later on, the line-tied conditions are applied using zero-velocities. For the rest of the variables, zero-derivative conditions are assumed.

\subsection{Mass loading and relaxation}\label{subsec:Mass loading}
In this work we do not consider the mechanisms of the prominence mass formation \citep[see, e. g.,][]{Xia:2011apj, Xia:2012apj, Luna:2012apj1, Xia:2016apj}. Instead of this, we place the prominence mass artificially in the dip of the flux rope, adding a source term in the continuity equation, as in \citet{Terradas:2013apj}. Thereby, in a defined region of the domain, the density increases in comparison to the surrounding corona while the temperature decreases, keeping the gas pressure constant. The corresponding source for the continuity equation is determined as follows
\begin{equation}\label{mass_source}
    S_{\rho}=\frac{\chi\rho_{0}}{t_{load}}\exp\left({-\frac{(x-x_p)^4}{\sigma_x^4}-\frac{(z-z_p)^4}{\sigma_z^4}}\right) \, ,
\end{equation}
where $\rho_{0}$ and $\chi$ are the background density and density contrast, respectively. The time parameter  $t_{load}=t_{4}-t_{3}=500\secs$ is associated with the loading rate where $t_{3}=5000\secs$ and $t_{4}=5500\secs$ determine the initial and final stages of the mass loading process. The spatial parameters $x_p=0 \Mm, z_p=8.4 \Mm$ are coordinates of the prominence center and $\sigma_x=3.6 \Mm, \sigma_z=2.4 \Mm$ are related to the half-size of the mass distribution region. In this study we have considered light and heavy prominences with $\chi=100$ and $\chi=200$.

The cancelation and the mass loading produce non-negligible vertical oscillations of the prominence structure. These oscillations are not desirable for our study. We have defined a relaxation process that removes these oscillations. During the relaxation process, the velocities in the domain are decreased due to artificial damping. At each iteration, the equations are updated with the new values of the velocities and the total energy. It is necessary to let this system evolve and relax for a long time ($4000\secs$ in our case). 

Figure \ref{fig:formation}(e) shows the density distribution at the beginning of the mass loading process.
The flux rope center drops down in order to find a new equilibrium position. The dense mass modifies the magnetic field lines making them more elongated downwards (see Fig. \ref{fig:formation}(e)). The magnetic forces increase as a result of lines stretching at the top and compression at the bottom. This way, magnetic forces compensate for the gravity increased due to the dense prominence mass. In the final stage of the relaxation, the prominence is close to equilibrium with velocities less than $2\kms$.
In the investigation of the prominence equilibrium in the 2.5D models, \citet{Hillier:2013apj} have shown that in the case of the absence of X-point, in the situation similar to ours, the prominence mass is supported by a combination of the magnetic tension force, magnetic pressure gradient, and the gas pressure gradient. 

In the final stage of the relaxation process, the prominence is located at a height of $7.4 \Mm$. The mass loading process produces a reconfiguration of the magnetic field and the plasma that produces a slight increase of the plasma-$\beta$ in the dipped part of the flux rope. However, this increase is small, and the low-$\beta$ situation is maintained.

\begin{figure*}[!ht]
\centering\includegraphics[width=0.9\textwidth]{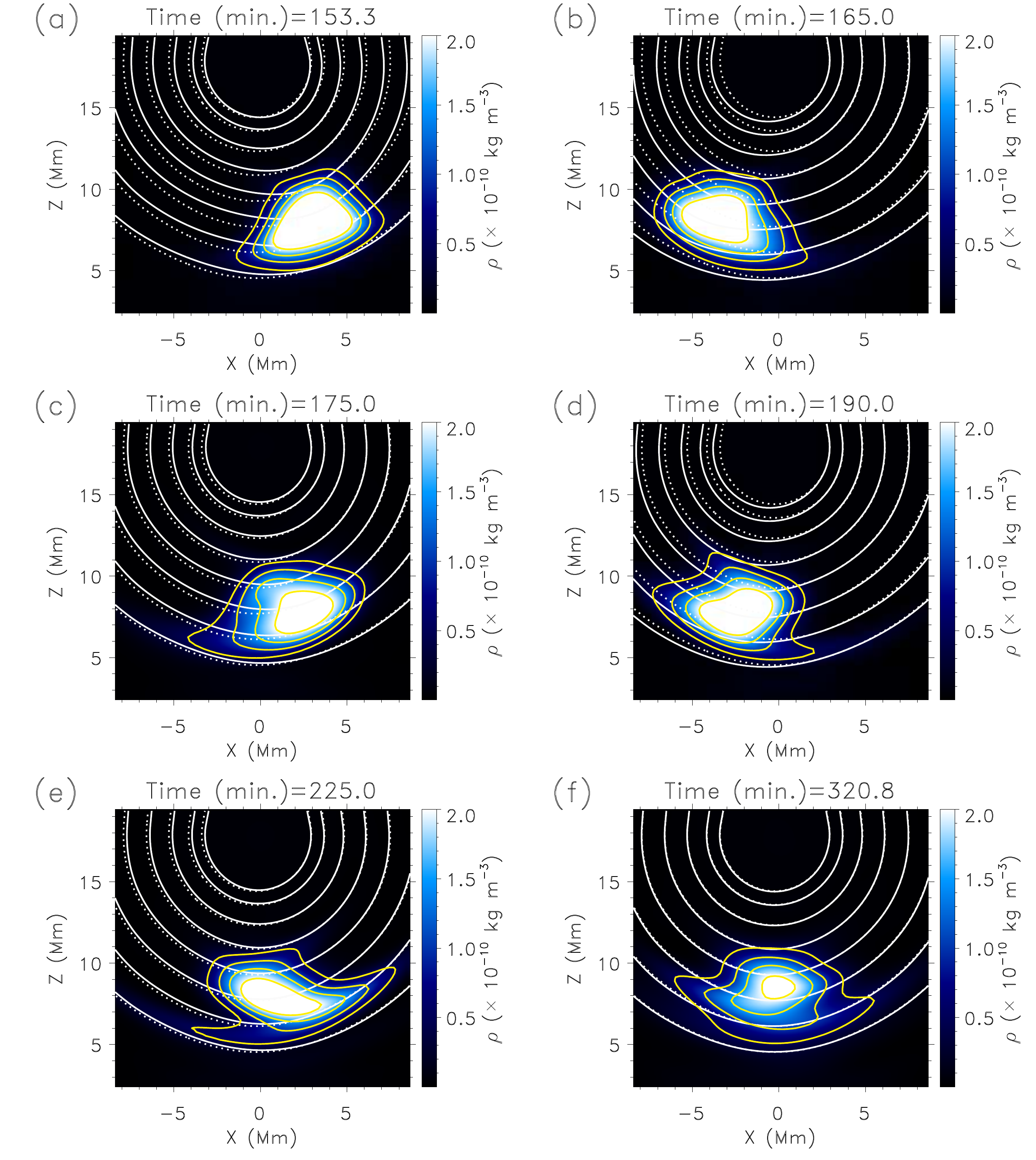}
\caption{Evolution of the density and the magnetic field after the horizontal perturbation. The figure shows the dipped part of the flux rope. Panel (a) demonstrates the motion of the prominence just after the perturbation; panels (b), (c) show the lagging of the bottom part from the rest of the prominence body, and panels (d), (e) demonstrate the zig-zag shape of the motions. Panel (f) shows the state of the prominence in the final stage of simulation. The dashed lines mark the magnetic field lines before the perturbation. The yellow lines represent the density isocontours.\label{fig:oscillations}}
\end{figure*}
\begin{figure}[!t]
\centering\includegraphics[width=0.5\textwidth]{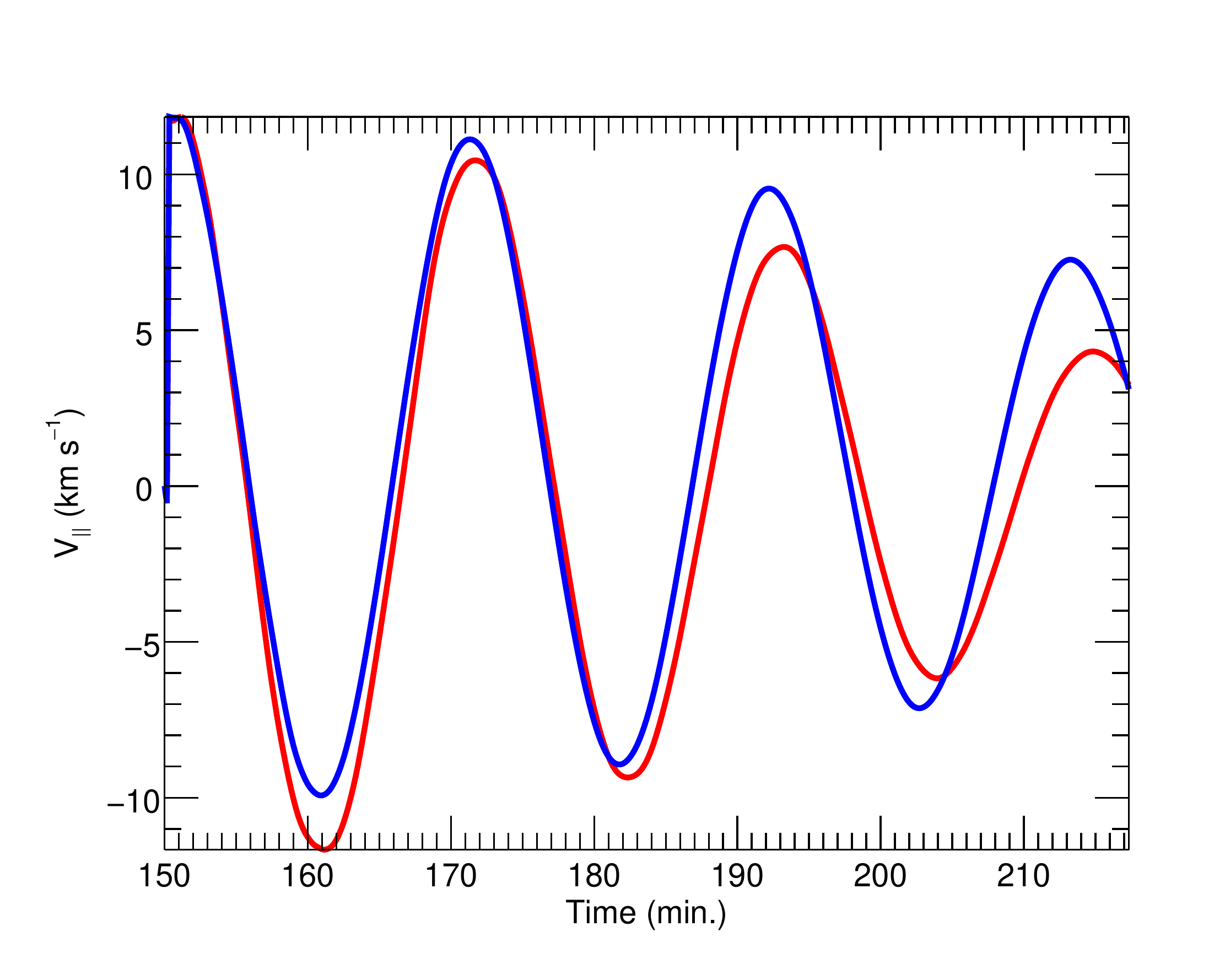}
\caption{The longitudinal velocity variations at the position of the center of mass of the prominence for the simulations with different resolutions. The red and blue lines are for the resolution of $240$ km and $60$ km, respectively.\label{fig:comparison}}
\end{figure}
\begin{figure*}[!ht]
     \centering\includegraphics[width=0.9\linewidth]{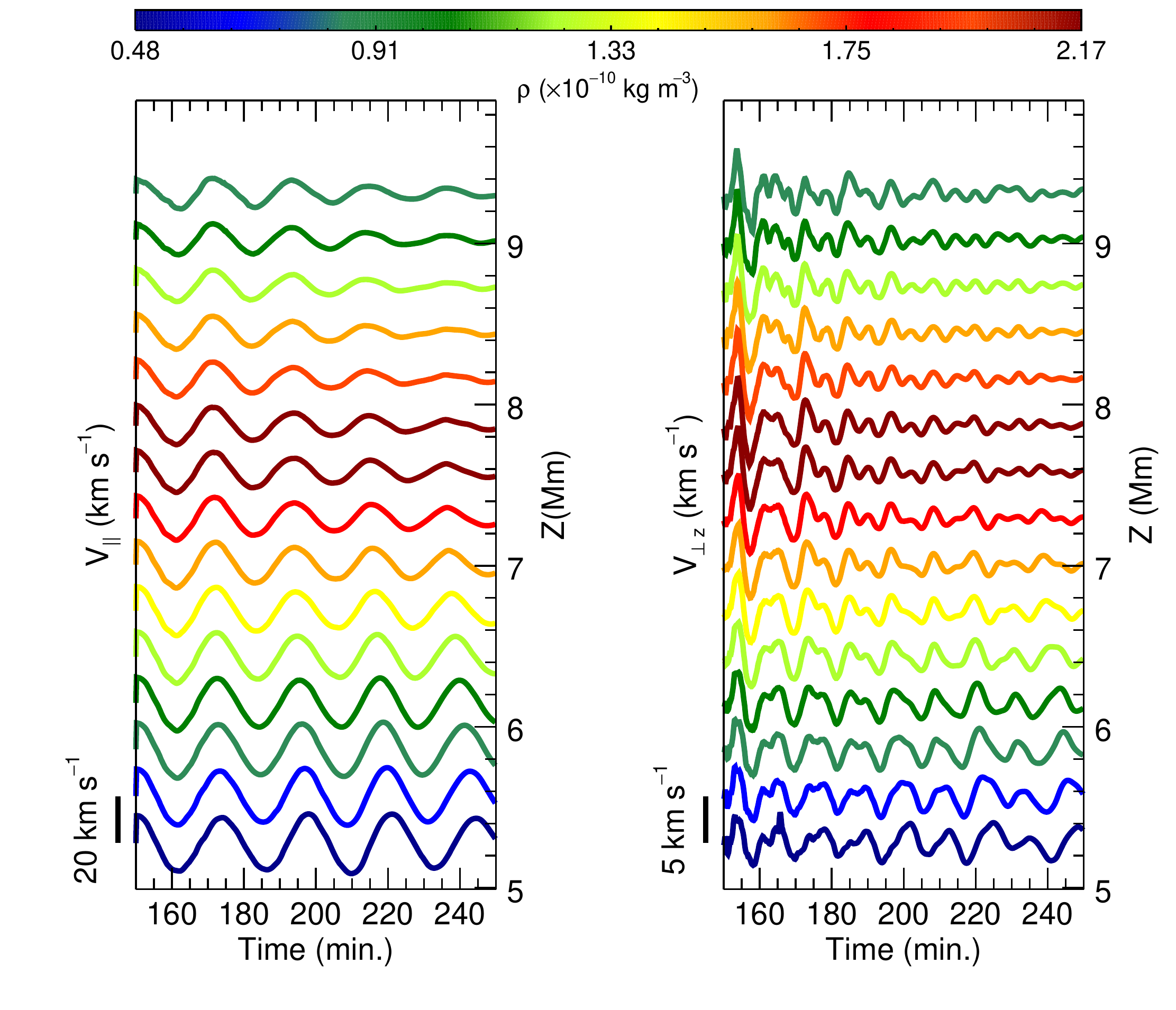}
     \caption{Temporal evolution of the longitudinal velocity (left panel) and the vertical component of the transverse velocity (right panel) after the horizontal perturbation. The prominence model used here has the shear angle of $\theta=45^{\circ}$, and the density contrast, $\chi=200$. Color lines indicate the velocity variations at the center of mass of different field lines. The colors mark the maximum initial density at the field lines according to the color bar scale. The right vertical axis indicates the height of the magnetic dips; the left vertical axis represents the scale of the maximum initial amplitude. \label{fig:signal1}}
\end{figure*}

\section{Oscillations due to the internal perturbations}\label{Internal}
After the flux rope formation, mass loading, and the relaxation process, we obtain a prominence close to a static equilibrium at $t=150.0\mins$, as we have described in Section \ref{Numerical}.
We first excite the system with internal perturbations in order to investigate the possible normal modes of different polarizations. This approach has already been used in several works \citep[see, e. g.,][]{Terradas:2013apj, Luna:2016apj, Zhou:2018apj} but for different configurations.
We consider four prominence configurations corresponding to a combination of the two shear angles, $\theta=45^\circ,60^\circ$, and the two maximum density contrasts, $\chi=100, 200$. Each system is perturbed in both vertical and horizontal directions.
The perturbations can be produced by introducing a pulse in some of the quantities. For example, \citet{Terradas:2013apj, Luna:2016apj, Zhou:2018apj} impose a perturbation exclusively in the velocity field, and the rest of the magnitudes remain unchanged. These kinds of perturbations are unphysical because the rest of the magnitudes should also be modified to fulfill the governing equations. However, the system rapidly accommodates them in a few time steps. In some situations, if the initial perturbation is large, the numerical simulation crashes.
We have adopted an alternative approach to perturb the system.
We have incorporated a source term for each of the MHD equations in the Mancha 3.0 code \citep{Felipe:2010apj}, similar to what we did for the mass loading (Sec. \ref{subsec:Mass loading}). The source term in the momentum equation represents an external force. This way, we produce self-consistent solutions. All the variables adjust to the source term following the MHD equations.
The system is perturbed in either vertical or horizontal directions by the source terms in the $z$- or $x$-projection of the equation of motion, correspondingly. The source has the following form
\begin{equation}\label{momentum_source}
S_{m}=\frac{\rho v_{pert}}{t_{pert}}\exp\left({-\frac{(x-x_{pert})^4}{\sigma_x^4}-\frac{(z-z_{pert})^4}{\sigma_z^4}}\right)\, ,
\end{equation}
where $t_{pert}=10\secs$  is the duration of the initial perturbation, $\sigma_{x}=\sigma_{z}=12\Mm$, $x_{pert}=0\Mm$ and $z_{pert}=7.4\Mm$ set the half-size and the $x$- and $z$-coordinates of the center of the perturbation, respectively. The maximum velocity of the perturbation,  $v_{pert}=20\kms$, is in the range of the observed amplitudes.

\subsection{Horizontal perturbation}\label{subsec:horizontal-perturbation}
Figure \ref{fig:oscillations} shows the time sequence of the density and the magnetic field evolution due to the horizontal perturbation for the case $(\theta,\chi)=(45^\circ,200)$. The three remaining cases are not shown because the temporal evolution is qualitatively similar. The solid lines represent the magnetic field, while the dashed lines mark the magnetic field lines just before the perturbation at $t=150.0\mins$ (at the end of the relaxation). The figure shows that the field lines change following the motion of the mass. However, the time-dependent changes in the shape of the magnetic field seem small. At the beginning, the prominence moves to the right and reaches its maximum displacement $3\mins$ after the perturbation (Fig. \ref{fig:oscillations}(a)). In this initial stage, the prominence moves mainly as a whole (Figs. \ref{fig:oscillations}(b) and \ref{fig:oscillations}(c)). However, the bottom part slightly delays from the rest of the prominence. Later, the oscillations become less synchronized, the phase difference increases, resulting in a zig-zag shape of the prominence (Figs. \ref{fig:oscillations}(d), \ref{fig:oscillations}(e)). In the recent work by \citet{Luna:2016apj}, the zig-zag shape appearance of LALOs was related to the differences in the period between the neighboring field lines. These phase shifts can make an impression of the vertical wave propagation. Such apparent wave motion has been called super-slow mode \citep{Kaneko:2015apj2, Raes:2017aa, Zapior:2019aa}.
\begin{figure*}[!ht]
     \centering
     \includegraphics[width=0.9\linewidth]{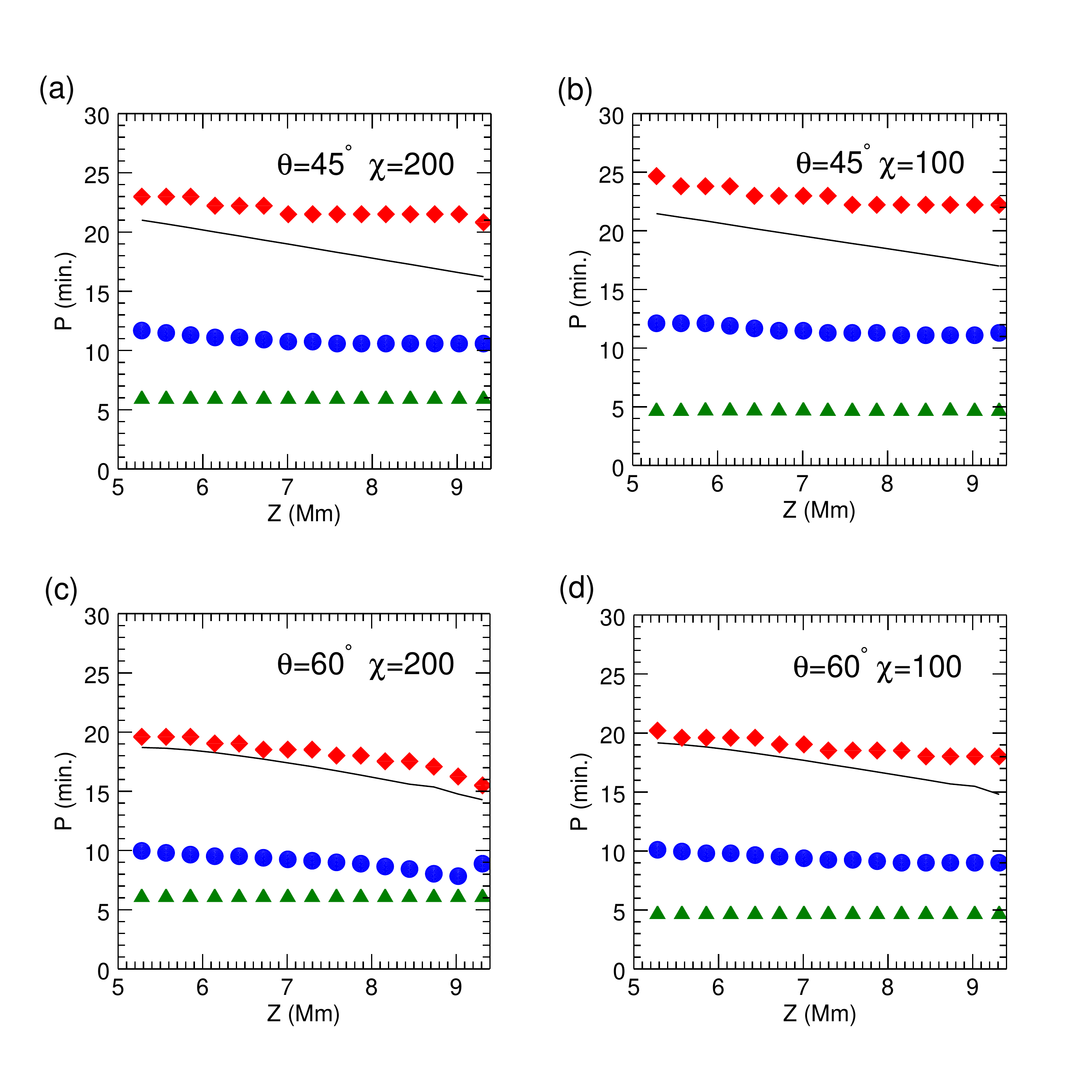}
     \caption{The oscillation periods as a function of height of the magnetic dips. Top panels correspond to the models with $\theta=45^{\circ}$, bottom panels correspond to the models with $\theta=60^{\circ}$. Left and right panels show the periods of the heavy and light prominence models, respectively. Red diamonds: longitudinal period, blue circles: vertical period due to the horizontal perturbation, green triangles: vertical period due to the vertical perturbation, solid line: period from the theoretical pendulum model.}\label{fig:periods}
\end{figure*}
\begin{table*}[!ht]
\centering   
\caption{The radii of curvature, the longitudinal and transverse vertical periods of the configurations with the different shear angle, $\theta$, and the density contrast, $\chi$.}              
\label{table1}      
\begin{tabular}{*{10}{c c c c c c c c c c}}
\hline\hline
\multirow{3}{*}{($\theta$, $\chi$)}  & \multicolumn{2}{c}{$X$-perturbation} & \multicolumn{2}{c}{$Z$-perturbation} & \multicolumn{2}{c}{External perturbation}\\
& $R_{c}$ & $P_{\parallel}$ &$P_{\parallel}$& $P_{v}$ & $P_{\parallel}$ & $P_{v}$ \\  
&(Mm) & (min.) & (min.) & (min.) & (min.) & (min.) \\ \hline
 $ (45^\circ,200) $ & $6.6-11.0$ & $20.8-23.0$& $12.8-13.9$ & $5.9$ & $18.5-23.8$ & $5.9$\\
 $ (45^\circ,100) $ & $7.2-11.5$ & $22.2-24.7$& $10.8-11.9$ & $4.7$ & - & -\\
 $ (60^\circ,200) $ & $5.1-8.7$ & $15.5-19.6$ & $14.8-16.3$ & $6.1$ & - & -\\
 $ (60^\circ,100) $ & $5.5-9.2$ & $18.0-20.2$ & $12.1-12.3$  & $4.7$ & - & -\\
 \hline
\end{tabular}
\end{table*}

From Figure \ref{fig:oscillations}(f), we see that the longitudinal oscillations are damped. 
This damping could be associated with the numerical dissipation.
To test the convergence of our simulations and verify to which extent the damping is caused by the numerical reasons, we have performed the simulations with a higher resolution ($60$ km).
Figure \ref{fig:comparison} shows the velocity of the center of mass of the prominence projected along the magnetic field for the simulations with 240 km and 60 km resolution.
This test shows that the damping in higher resolution simulation is weaker, indicating that the damping has a numerical contribution. It would be necessary to perform a dedicated study to understand this damping further by considering consecutively higher spatial resolutions. This can be a subject of future work. For both resolutions considered here, the periodicity is similar. We then conclude that the system has converged.
An additional effect of the numerical diffusivity is a slight reduction of the prominence density with time. The high-resolution simulation shows a smaller reduction of the density. However, this density reduction does not seem to affect the prominence dynamics.
Taking into account these results, it is enough for our purposes to consider a spatial resolution of  $240$ km.

As we see in Figure \ref{fig:oscillations}, different parts of the structure oscillate differently, with different periods and phases. In addition, these motions are essentially nonlinear, and the plasma is a subject to important advection. The frozen-in condition applies in our fluid because we consider it to be perfectly conducting. Thus, it is necessary to advect every field line considered in order to catch the longitudinal and the transverse motions of the plasma. We are interested in the cold plasma of the prominence, where the $xz$-projection of the field lines are more or less ellipses (Fig. \ref{fig:formation}(f)). We have realized that the top part of these ellipses is approximately unperturbed.  Thus, any field line starting at the top of the flux rope follows the plasma motion, simplifying the study of the oscillations. Following \citet{Luna:2016apj}, we calculate the longitudinal and transverse velocities at the positions of the center of mass of each field line. We have selected 15 field lines of the flux rope with dips located at heights between $5\Mm$ to $10\Mm$.

The temporal evolution of the longitudinal velocity for each of the selected field lines is shown at the left panel of Figure \ref{fig:signal1}. The velocity amplitude of these oscillations depends on the field line considered because of the shape of the triggering (Eq. \eqref{momentum_source}). The initial velocity is around $20\kms$ for the field lines centered at the cold plasma around $z=7.5\Mm$ and is smaller for the rest. For the field lines with $z>6\Mm$, the oscillations show damping that depends on the height. However, for the field lines with $z<6\Mm$, the amplitude is constant, or it slightly amplifies at the half time shown in the figure. 
From the edge of the flux rope to its center, the magnetic field lines become more twisted. Thus, we conclude that oscillations are strongly damped at places, where the lines have large curvature.
In contrast to \citet{Luna:2016apj}, the phase differences between the lines are small. However, the combination of these small phase shifts and the different damping of the different layers produces the zig-zag shape appearance.

The horizontal perturbation excites not only the longitudinal oscillations but also the transverse ones, as we see at the right panel of Figure \ref{fig:signal1}. This figure shows that the initial vertical velocity is zero at all the heights of the structure. However, immediately, the system starts to oscillate with the amplitudes of approximately $5\kms$. Similarly to the longitudinal oscillations, for heights $z>6\Mm$, the motions are clearly damped, whereas for $z<6\Mm$, the plasma keeps oscillating.

We repeat the previous analysis for all four cases obtaining the temporal evolution at the different layers of the different prominence configurations. We have analyzed the periods of oscillations at every field line considered, and in the two polarization directions. The results for all four $(\theta,\chi)$ configurations are plotted in Figure \ref{fig:periods}. The red diamond symbols show the period of the longitudinal oscillations as a function of the height of the dips. In all four cases, we can see that the period tends to decrease with height. The difference between the minimum and the maximum value is $2-3\mins$ (see Table \ref{table1}). The longitudinal period slightly changes for different field lines. As a result, we have the zig-zag shape of the prominence after several cycles of oscillations (Fig. \ref{fig:oscillations}(e)). The vertical period due to the horizontal perturbation is plotted with the blue circle symbols in Figure \ref{fig:periods}. The vertical oscillations have periods that are half of the longitudinal periods. The behavior with height is identical to that of the longitudinal oscillation. This result indicates that the vertical motion is the back-reaction to the longitudinal oscillations, as we have already seen in Figure \ref{fig:signal1}. For example, when the prominence mass passes through the dip, it pushes the structure downwards. This happens twice for each period explaining why the vertical oscillations have half of the period of the longitudinal one.

We compute the radii of curvature averaged around the bottom of the dip to compare our results with the pendulum model. The magnetic field evolves due to the motion of the prominence plasma. Because of that, we use the time-averaged values of the radii of curvature, $R_{c}$. It allows us to calculate the theoretical period of the longitudinal oscillations.  We obtain the pendulum period shown as a solid line in Figure \ref{fig:periods} by using equation, $P=2\pi \sqrt{R_{c}/g}$, from \citet{Luna:2012apj} where $g=273.94\ \mathrm{m\ s^{-2}}$ is the solar gravitational acceleration. For $\theta=45^\circ$ the periods obtained in the numerical experiment show a good agreement with the theoretical period for $z<7.5\Mm$ and larger discrepancy for larger heights. In contrast, for $\theta=60^\circ$ the agreement is much better in all the spatial domain. We conclude that, for this type of oscillation, the restoring force is the projected solar gravity. The remaining difference between the numerical experiment and the theory of about $7\%-15\%$ can be related to the local deformation of the magnetic field lines due to the motion of the heavy mass. \citet{Zhang:2019apj} and \citet{Zhou:2018apj} have shown by 2D and 3D simulations, respectively, that the pendulum model might lead to a large error in determining the radius of curvature using longitudinal periods when the gravity to Lorentz force ratio is close to unity. The authors defined a dimensionless parameter to quantify this ratio as ${\delta=2gL_{p}/V_{A}^{2}}$ being $L_{p}\cong10 \Mm$ a typical dimension of the prominence and the local Alfv\'{e}n speed, $V_{A}\cong100\kms$, in the case of $\chi=200$. In our situation, this parameter is close to $\delta=0.6$, indicating that the deformation of the field lines by the gravity could have some moderate effect on the longitudinal oscillations.

A comparison between the different simulations in Figure \ref{fig:periods} shows that a change of the shear angle or the density contrast does not lead to a large difference in the longitudinal period. Comparing the heavy and light prominence models in Figure \ref{fig:periods}, we can see that the longitudinal period is only $1-2\mins$ longer in the case of the lower density contrast, $\chi=100$. Since we have concluded that the projected solar gravity is the main restoring force, the period should essentially depend only on the radius of curvature. It means that the radii of curvature are slightly larger, where prominence is less massive. As was mentioned before, the loaded cold mass deforms the magnetic field lines decreasing the radius of curvature (Fig. \ref{fig:formation}(f)). It is expected that heavier mass affects the shape of the field lines more, as a consequence, decreasing the longitudinal period.

Upper and bottom panels of Figure \ref{fig:periods} compare oscillations for the models with $\theta=45^{\circ}$ and $\theta=60^{\circ}$. The longitudinal period does not show a large difference in these two cases. The difference of $3-4\mins$ can be related to the projection effect. The more sheared structure has a smaller radius of curvature in the 2D projection (Table \ref{table1}). As a result, the longitudinal period is also slightly shorter.

In order to estimate the damping time, the signal of the longitudinal velocity at the position of the center of mass of the prominence is fitted by the decayed sinusoidal function $v=v_{0}e^{-t/{\tau_{D}}}\sin{(2\pi t/P+\phi)}$. We computed the damping time for the case shown in Figure \ref{fig:signal1}. For the lines at the bottom of the prominence, the damping time, $\tau_{D}=100\mins$, while at the top, it is only $60\mins$. Thus, the damping time shows a decrease with height, in agreement with the visual impression from the left panel of Figure \ref{fig:signal1}. We have repeated the analysis for each of the models and obtained a similar behavior of damping with height.

\begin{figure*}[!h]
\includegraphics[width=0.9\textwidth]{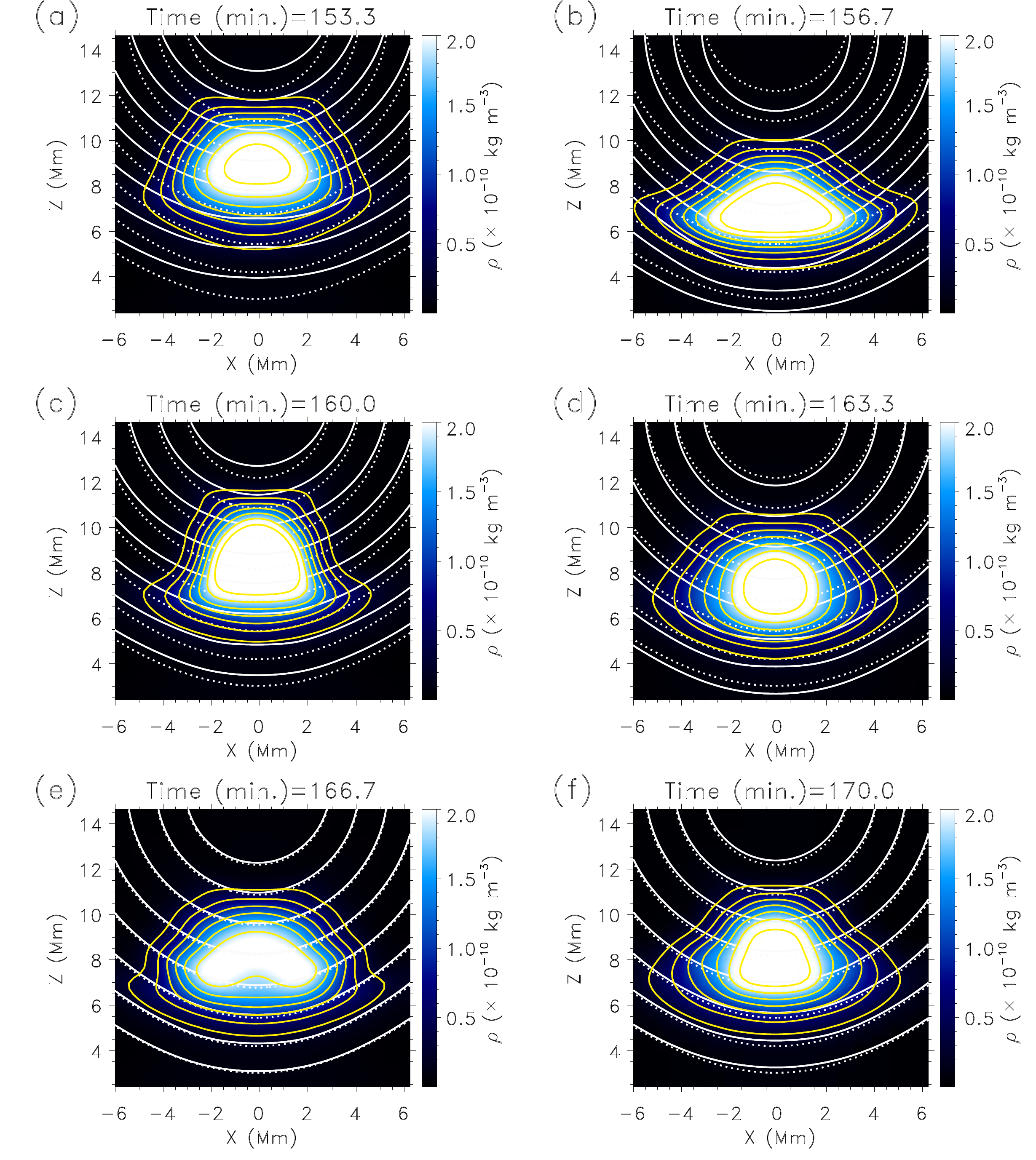}
\centering
\caption{Evolution of the density and the magnetic field after the vertical perturbation. The panels (b)-(f) demonstrate the longitudinal motions at the bottom of the prominence due to compression and rarefaction. The dashed lines mark the magnetic field lines before the perturbation. The yellow lines represent the density isocontours. \label{fig:vertical}}
\end{figure*}
\begin{figure*}[!ht]
     \centering
     \includegraphics[width=0.9\linewidth]{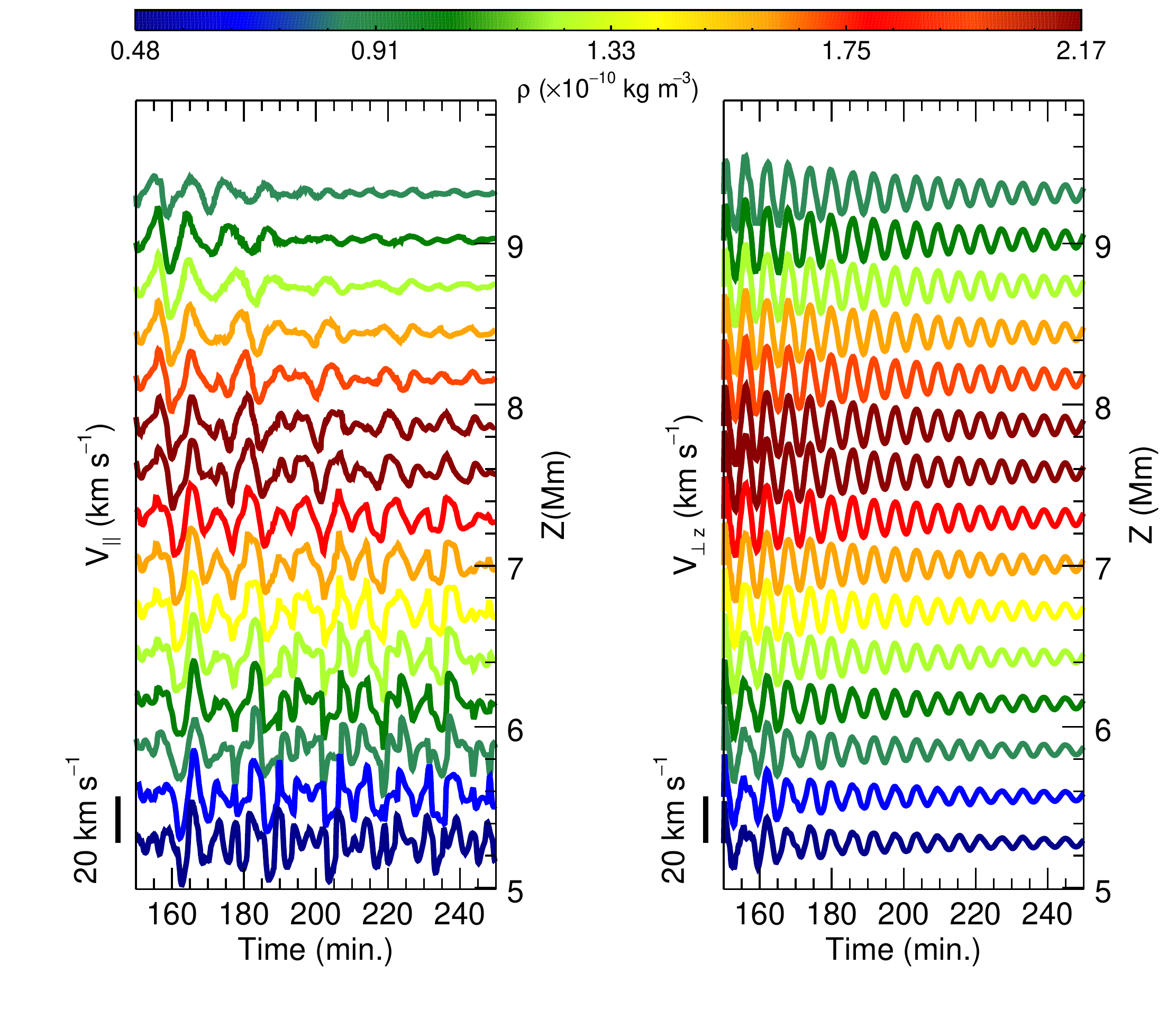}
     \caption{Temporal evolution of the longitudinal velocity (left panel) and the vertical component of the transverse velocity (right panel) after the vertical perturbation. The prominence model used here has the shear angle $\theta=45\,^{\circ}$, and the density contrast, $\chi=200$. Color lines indicate the velocity variations at the center of mass (right) or the certain distance from the center of mass (left) at each field line. The colors mark the maximum initial density at the field line according to the color bar scale. The right vertical axis indicates the height of the magnetic dips; the left vertical axis represents the scale of the maximum initial amplitude.}\label{fig:signal2}
\end{figure*}
\begin{figure*}[!ht]
     \centering
     \includegraphics[width=0.9\linewidth]{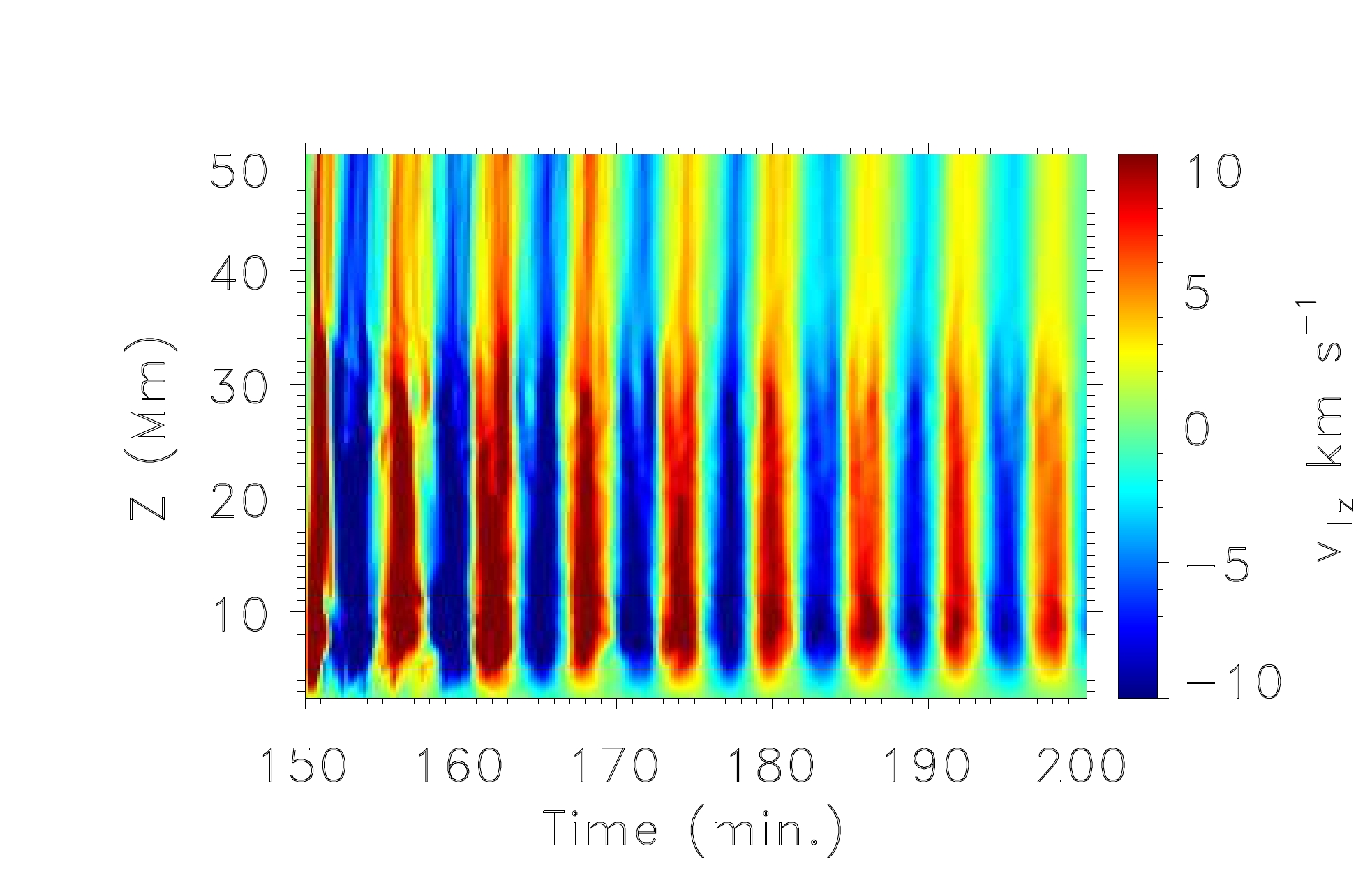}
     \caption{The time-distance diagram of the vertical component of the transverse velocity along $z$ - axis at $x=0$, in the model with the shear angle $\theta=45\,^{\circ}$, and the density contrast, $\chi=200$. Black lines mark a region where the prominence mass is located.}\label{fig:time_distance}
\end{figure*}

\subsection{Vertical perturbation}\label{subsec:vertical-perturbation}
We perturb the prominence along the vertical direction by applying the source term (Eq. \eqref{momentum_source}) to the $z$-component of the equation of motion. The temporal evolution of the prominence mass just after the vertical perturbation is shown in Figure \ref{fig:vertical}. Figure \ref{fig:vertical}(a) shows the vertical upward motion of the prominence right after the perturbation. We see how the field follows the plasma motion due to the frozen-in condition, and the dips flatten. In Figure \ref{fig:vertical}(b), the direction of the motion has changed. The motion is downward, and the magnetic field is bent in comparison to the initial configuration. In addition to the vertical motion, the prominence is a subject to the longitudinal motions. These longitudinal motions are due to the compression and the rarefaction of the plasma. In Figure \ref{fig:vertical}(b), we can see clearly how the prominence expands symmetrically with respect to the $x=0$ axis along the field lines. In Figures \ref{fig:vertical}(d) and \ref{fig:vertical}(f), prominence shows symmetric longitudinal motions. This motion is more noticeable as the difference between the behavior at the bottom and at the upper prominence layer. In Figure  \ref{fig:vertical}(f), it is seen that the bottom part is more expanded than the upper part.

The behavior of the different components of the velocity at several prominence layers is shown in Figure \ref{fig:signal2}. The right panel of Figure \ref{fig:signal2} shows the vertical oscillations at the center of mass of each field line, as in Section \ref{subsec:horizontal-perturbation}. However, due to the symmetry of the vertical oscillations, the longitudinal velocity is zero at the center of mass of each line. For this reason, we measured the longitudinal velocity outside of the center close to the position of the maximum velocity. Figure \ref{fig:signal2}(left) shows the temporal evolution of the longitudinal velocity at the selected field lines. We can see that the initial amplitude is zero, but rapidly the longitudinal oscillations start.  Figure \ref{fig:vertical}(b) shows the expansion of the prominence body when it moves downwards. The amplitudes increase, reaching the value of $20\kms$ at some lines. A very interesting feature of this numerical experiment that a vertical perturbation also produces an important horizontal oscillation of the cool plasma. It is unclear if, in the first stages of the oscillation, there is a mode coupling between vertical and horizontal polarization modes or, in contrast, it is just the transition phase of the establishment of modes, where the oscillation is composed of vertical and horizontal compression and rarefaction movements.
Generally, the energy transfer could be related to the fast-to-slow mode conversion in the region where plasma-$\beta$ is close to unity \citep{Schunker:2006mnras, Chen:2016solphys} or to the resonant absorption in the Alfv\'{e}n continuum at the prominence-corona transition region (PCTR) \citep{Goossens:2014apj, Antolin:2015apj}. Since we have plasma-$\beta$ lower than unity everywhere in the numerical domain, the mechanism of the fast-to-slow mode conversion is not taking place.
The resonant absorption could still be responsible for the energy transfer and the damping of the global prominence modes. If this mechanism is in action, the global mode of the prominence is transferred to the local modes in the magnetic surfaces of the PCTR. We have studied the distribution of the kinetic energy with time, and we have not detected any transfer of energy to magnetic surfaces of the structure.  Unlike our case, the energy of transverse oscillations was shown to be transferred to a thin region at the PCTR in 3D MHD simulations by \citet{Terradas:2016apj}. The non-existence of this phenomenon in our case could be associated with limitations of the 2.5D modeling. The resonant absorption in 3D flux rope structures will be the subject of a future study.

After 2 minutes, the longitudinal motions seem to be established. However, the oscillations are very irregular with a complex pattern.
These longitudinal motions are associated with compression and rarefaction, as is shown in Figures \ref{fig:vertical}(a)-\ref{fig:vertical}(f), rather than related to the projected solar gravity.
The periods of these longitudinal modes (see Table \ref{table1}) are shorter than those of the pendulum-like motions of Section \ref{subsec:horizontal-perturbation}. Also, these periods are not identical or the double of the vertical periods indicating that the longitudinal modes are not related to the vertical modes. Altogether, this may indicate that the vertical perturbation also excites overtones of the structure with antisymmetric velocity profiles, as studied in \citet{Joarder:1992aa, Oliver:1993co, Luna:2012apj2}.
 The temporal evolution of the vertical velocity is shown at the right panel of Figure \ref{fig:signal2}. The vertical oscillations in the different layers of the structure show a very coherent motion, with oscillations almost in phase at all heights. The periods are very uniform with values shown in Table \ref{table1} and in Figure \ref{fig:periods}. Also, in all the field lines, the attenuation is moderate and very uniform along the vertical direction. As we have explained previously, there is no resonant absorption or fast-to-slow mode conversion, and they can be discarded as possible damping mechanisms. However, this attenuation can be explained by numerical diffusivity or by the wave leakage. Figure \ref{fig:time_distance} shows a time-distance diagram of the vertical velocity along a vertical cut at the center of the structure. The red and blue colors correspond to the upward and downward motions, respectively. The prominence is located between $5\Mm$ to $10\Mm$. As we can see from this figure, strong vertical motion appears after the perturbation at heights above the prominence, and it extends even higher. The vertical bands have a slight inclination. This inclination reveals that there is a propagation of the velocity perturbations in the form of waves. The inclination is consistent with a velocity of $1000 \kms$. This means that some energy leaves the system during this time due to the wave leakage. Taking into account the attenuation shown at the right panel of Figure \ref{fig:signal2}, we conclude that this wave leakage contributes to the damping of the vertical oscillations. We calculate the damping time, as described in Section \ref{subsec:horizontal-perturbation}. The damping time is about of $60\mins$ in almost all the lines and is slightly shorter at the bottom (of about $50\mins$).

In several works \citep{Hyder:1966zap, Kleczek:1968solphys, Luna:2016apj, Zhou:2018apj}, the authors have concluded that the restoring force for the transverse oscillations has a magnetic origin. We have calculated the magnetic forces and their relative variations as it was done by \citet{Zhou:2018apj}. The variations of the magnetic pressure gradient exceed the variations of the magnetic tension force, for all considered configurations. It indicates that in our model, the magnetic pressure gradient is the main restoring force for the vertical oscillations.

We have obtained the period of the vertical oscillations as in Section \ref{subsec:horizontal-perturbation}(Table \ref{table1}).The vertical period is plotted with green triangles as a function of the height of the dips in Figure \ref{fig:periods}. The oscillatory period remains constant at different layers of the prominence for all models. This indicates that the vertical motion is associated with global normal mode \citep{Terradas:2013apj, Luna:2016apj}. From Figure \ref{fig:periods}, we see that the transverse periods are much shorter than longitudinal ones, similar to as revealed by observations \citep[see review][Table 1]{Tripathi:2009ssr}. Same as the longitudinal period, the vertical period does not change much for different models, with an average difference of around $1 \mathrm{\, minute}$ (see Table \ref{table1}). Part of the variation of the vertical period for different density contrast can be due to the inertia effect. For the more massive prominence, with the density contrast $\chi=200$, the vertical period is slightly longer (Figs. \ref{fig:periods}(a) and \ref{fig:periods}(c)).
Comparing the top and bottom panels of Figure \ref{fig:periods}, we can see that the vertical period seems to be almost insensitive to the shear angle variation.

\begin{figure*}[!ht]
\includegraphics[width=0.8\textwidth]{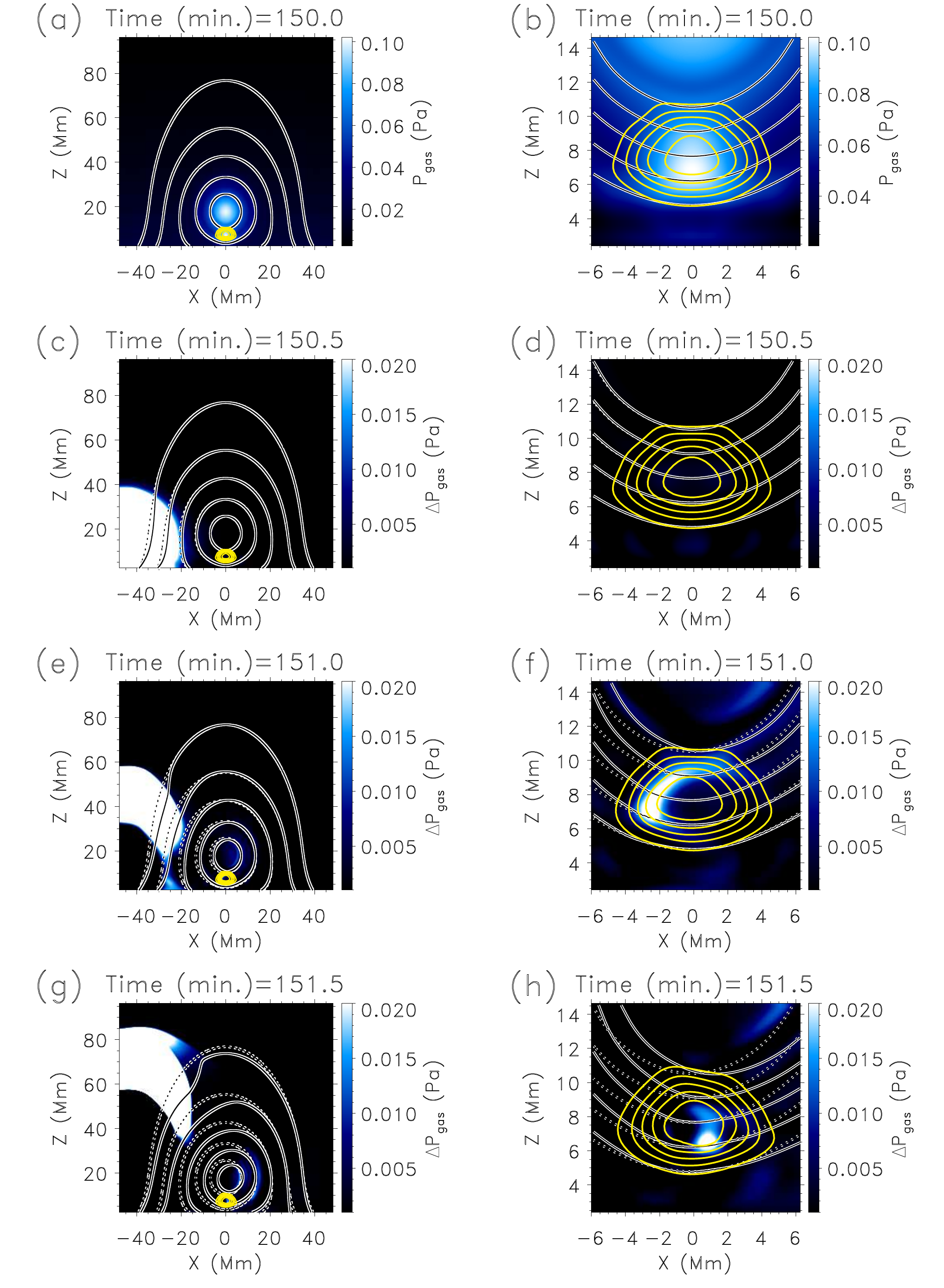}
\centering
\caption{ Evolution of the magnetic field and the gas pressure during the wavefront propagation. Panels (a), (b): the initial gas pressure distribution; panels (c)-(h): the running difference of the two consecutive snapshots. The left-hand side panels show the same as the right-hand side panels but focused at the location around the prominence. The dashed lines mark the magnetic field lines before the perturbation. The yellow lines represent the density isocontours.\label{fig:external}}
\end{figure*}
\begin{figure*}[!ht]
     \centering
     \includegraphics[width=0.9\linewidth]{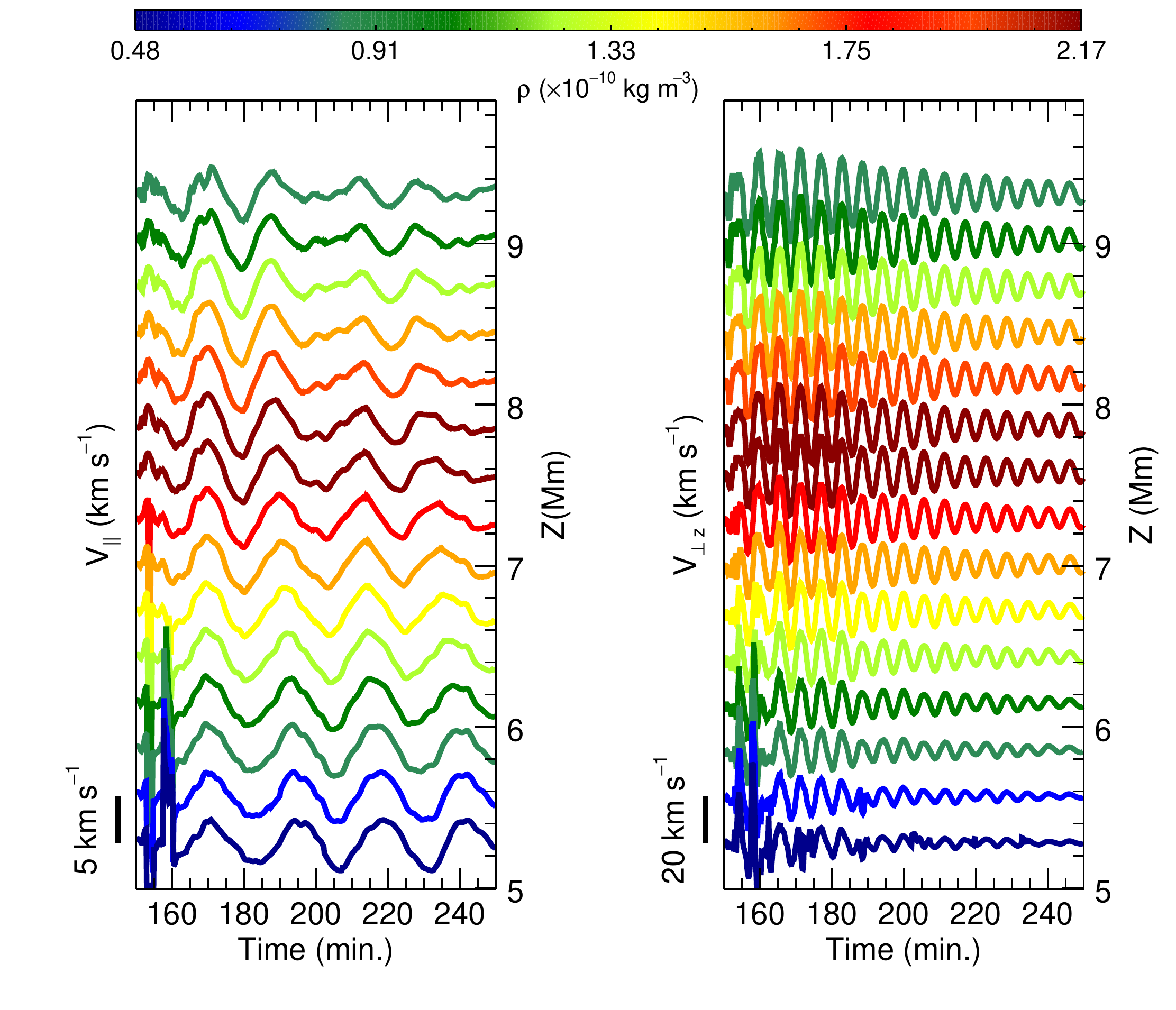}
     \caption{Temporal evolution of the longitudinal velocity (left panel) and the vertical component of the transverse velocity (right panel) after the wavefront propagation. The prominence model used here has the shear angle of $\theta=45^{\circ}$, and the density contrast, $\chi=200$. Color lines indicate the velocity variations at the center of mass of different field lines. The colors mark the maximum initial density at the field lines according to the color bar scale. The right vertical axis indicates the height of the magnetic dips; the left vertical axis represents the scale of the maximum initial amplitude. \label{fig:signal3}}
\end{figure*}

\section{Oscillations due to the external perturbation}\label{External}

LAOs are usually triggered by external perturbations as Moreton or EIT waves. It is crucial to understand how the external perturbation interacts with the flux rope structure and produces  oscillations of the cold prominence. We model this scenario by using the perturbation that arrives from outside. In order to produce strong perturbation comparable to the realistic energetic events, we use the source term of the total energy equation instead of perturbation of pressure or velocity. In this case, we add energy to the system, all variables are adjusted following the MHD equations, and we have a self-consistent perturbation (see the discussion on Sec. \ref{subsec:horizontal-perturbation}). We place the perturbation at some distance from the prominence center as follows:
\begin{equation}\label{energy_source}
S_{e}=\frac{\alpha}{t_{pert}}\exp\left({-\frac{(x-x_{pert})^2}{\sigma_x^2}-\frac{(z-z_{pert})^2}{\sigma_z^2}}\right)\, ,
\end{equation}
where the parameter $\alpha=2\ \mathrm{J\ m^{-3}}$. In solar flares temperature and density can be up to $10^{7}$ K - $10^{8}$ K and $10^{16}\ \mathrm{m^{-3}}$ that corresponds to the parameter $\alpha$ from $2\ \mathrm{J\ m^{-3}}$ to $20\ \mathrm{J\ m^{-3}}$. With this parameter, we produce an initial corresponding velocity around of $360\kms$. This value is typical for EIT waves that are usually associated with flares or coronal mass ejections \citep{Moses:1997solphys, Thompson:1998grl}. In turn, $t_{pert}=50\secs$, $x_{pert}=-45.6 \Mm$, $z_{pert}=12 \Mm$, $\sigma_{x}=\sigma_{z}=12  \Mm$ are the same parameters as in Equation \eqref{momentum_source}.

Our aim is to study how the wavefront propagates and interacts with the flux rope. The source of the wave should be placed at a certain distance from the flux rope. We consider a periodic system, virtually composed of an infinite number of arcades separated by a region of the vertical field. In our domain, we have not only the main arcade but also the neighboring arcades.
If the perturbation would be located far (in the area of the neighboring arcade), the wavefront would propagate along the field lines to the boundary, and most energy would remain trapped in the neighboring arcade with only a small part of the energy reaching the prominence. 
To avoid this behavior, we have chosen the initial position of the perturbation between the arcades, where the magnetic field is predominantly vertical. 
The temporal evolution of the system is shown in Figure \ref{fig:external} (left panels). The right-hand side panels show the same, but with a more detailed view around the cold prominence. Figures \ref{fig:external}(a) and \ref{fig:external}(b) show the initial distribution of the gas pressure. We can see a high pressure inside the flux rope in comparison with the environment. In this region, pressure increases during the flux rope formation process. 
In order to visualize the propagation of the perturbation, we have computed the running differences that consist by subtracting two consecutive snapshots separated by 30 seconds (Figs. \ref{fig:external}(c)-\ref{fig:external}(h)).
As we can see in Figures \ref{fig:external}(c) and \ref{fig:external}(d), just after the source term activation, the external perturbation propagates mainly vertically, producing significant changes in the magnetic field. The field lines are displaced to the right with respect to the equilibrium structure (dashed lines). 
The external gas pressure perturbation is unable to cross the flux rope and does not reach the internal region of the prominence.
However, the compression of the lines creates a gas pressure excess inside the flux rope near the cold prominence plasma (Figs. \ref{fig:external}(e) and \ref{fig:external}(f)). From Figures \ref{fig:external}(e) and \ref{fig:external}(f), we see that pressure perturbation inside the flux rope propagates crossing the cold prominence, but the prominence is not deformed or displaced by this perturbation. 
Figure \ref{fig:external}(g) shows that the external wavefront continues propagating along the arcade field lines producing significant perturbations of the magnetic field. In Figure \ref{fig:external}(h), we see that the internal gas pressure pulse has crossed the prominence, slightly deforming it. After this transient, an oscillation is established. From this temporal evolution, we see that the magnetic field is responsible for the movement of the prominence. The external perturbation produces an Alfv\'enic disturbance that displaces the flux rope, and then the cold prominence mass moves. In this sense, the motion of the prominence is secondary because the perturbation firstly perturbs the flux rope.

\citet{Chen:2016solphys} have studied the interaction between the fast-mode shock wave and the arcade magnetic field structure using 2D MHD simulations of the periodic arcade magnetic field. The system did not include the prominence mass. The authors have concluded that if a wave vector is parallel to the magnetic field and the wave propagates in the region where plasma-$\beta$ is close to unity, the fast-to-slow mode conversion happens \citep[see][Fig. 7]{Chen:2016solphys}.  In the present case, the shock wave propagates in the low-$\beta$ environment, and we do not observe the fast-to-slow mode conversion.

As in the previous sections, we have obtained the longitudinal and the vertical velocities at the field lines center of mass, between $z=5$ Mm to $10$ Mm. 
From both panels of Figure \ref{fig:signal3}, we can see a complex dynamical behavior. In the longitudinal velocity, there are no clear signatures of oscillations in the first 10 minutes after the perturbation. In the bottom part of the structure, there are strong peaks associated with the external disturbance that propagates along the circular magnetic field lines of the flux rope. After some time, it reaches the bottom part of the flux rope, producing these spikes.
The oscillations have an initial amplitude of about $5\kms$. The period of the longitudinal oscillation slightly decreases with height similarly as at the left panel of Figure \ref{fig:signal1}. 
The vertical oscillations are established after the first 5 minutes of the perturbation (see Fig. \ref{fig:signal3} right panel). The amplitude of the vertical oscillations is around $20\kms$. We can see that the signal reaches the maximum amplitude after several cycles of oscillations. The motions are coherent in the different prominence layers. 
 The damping of the signal is similar to the situation shown at the right panel of Figure \ref{fig:signal2}.  Therefore, as in the case of the vertical perturbation, this attenuation can be associated with the wave leakage (see Sec. \ref{subsec:vertical-perturbation}). The periods of oscillations shown in Table \ref{table1} are in agreement with internal modes. 
From this numerical experiment, we conclude that external perturbation can excite both longitudinal and vertical oscillations. Even if the vertical oscillations are excited with larger amplitudes, they are also damped faster than the longitudinal modes. In this situation, the oscillation is mainly vertical just after the triggering, but the final phase contains mainly longitudinal oscillations.

\section{Summary and Conclusions}\label{Conclusions}

To study the physics of the LAOs, we have performed 2.5D numerical simulations using the mechanism of \citet{vanBallegooijen:1989apj} for the flux rope formation. Once the flux rope was formed, the cold mass is artificially loaded in the dipped part of the magnetic structure. After the relaxation process, we have applied the horizontal and vertical perturbations at the prominence position and have studied internal modes of different polarizations. We have also considered the case of the external LAOs triggering using the disturbance placed out of the flux rope. 

The horizontal perturbation triggers both longitudinal and transverse vertical oscillations. The period of the longitudinal oscillations, obtained numerically, shows a good agreement with the pendulum period. This confirms that, under the prominence conditions, gravity is the dominant restoring force for the longitudinal oscillations. The damping of the longitudinal motions is mostly related to the numerical dissipation. The vertical oscillations are associated with the backreaction of the magnetic field to the longitudinal motions.

The vertical perturbation drives vertical motions together with the longitudinal component, associated with compression and rarefaction. The period of the vertical oscillations remains constant with height. This indicates that the vertical disturbance triggers global normal mode of the structure. The oscillations show attenuation. We have found indications that the wave leakage can be responsible for the damping of the vertical oscillations.

We have considered oscillations of different polarizations using models with two values of the shear angle and the density contrast. We have compared the periods of the different prominence configurations. The periods of the longitudinal and vertical oscillations show weak dependence on these parameters of the prominence structure. In the case of the longitudinal period, the small difference in the period is related to the variations of the radius of curvature. In contrast, the vertical periods are slightly different for the heavy and light prominences due to the inertia effect.

We have investigated the triggering mechanism by external disturbance choosing the location of the source at a certain distance from the flux rope. The external perturbation reaches the flux rope displacing and deforming the magnetic field lines. The flux rope displacement creates perturbation inside the prominence, and, as a consequence, the prominence is also displaced from the equilibrium position. The signal analysis related to the different field lines has shown that the plasma motions have a complex character, and external disturbance drives both longitudinal and transverse vertical LAOs. 

Using the time-dependent numerical simulations, we have demonstrated that the period of the longitudinal oscillations is in a good agreement with the pendulum model, and the period of the vertical oscillations only slightly depends on the prominence density and the shear angle of the magnetic field. We have also proposed the model of the external LAOs excitation.  The simulation has shown that the external disturbance excites the normal prominence modes of the different polarizations. 

In the future, it will be necessary to investigate the oscillatory properties, using 3D prominence models. Firstly, in a 3D model, the flux rope can be anchored at the footpoints, which allows the magnetic tension to make a significant contribution to the restoring force. Therefore, including the perpendicular direction can significantly change the oscillatory properties of this model. Secondly, 3D simulations can allow us to study the external modes triggering in more detail. It is possible to reproduce the scenario when a wave arrives from different directions with respect to the prominence spine. For future work, it is essential to consider more realistic external disturbances such as an energy release due to the magnetic reconnection, or the energetic wave associated with the prominence eruption.

\begin{acknowledgements}  V. Liakh acknowledges the support of the Instituto de Astrof\'{\i}sica de Canarias via an Astrophysicist Resident fellowship. M. L. acknowledges the support by the Spanish Ministry of Economy and Competitiveness (MINECO) through Severo Ochoa Program MINECO SEV-2015-0548. E. K. thanks the support by the European Research Council through the Consolidator Grant ERC-
2017-CoG-771310-PI2FA and by the Spanish Ministry of Economy, Industry and Competitiveness through the
grant PGC2018-095832-B-I00 is acknowledged. V. Liakh, M. Luna, and E. Khomenko thankfully acknowledge the technical expertise and assistance provided by the Spanish Supercomputing Network (Red Espa\~{n}ola de Supercomputac\'{\i}on), as well as the computer resources used: the LaPalma Supercomputer, located at the Instituto de Astrof\'{\i}sica de Canarias. V. L. also acknowledges the support from ISSI \url{www.issibern.ch} to team 413 and team 431 on "Large-Amplitude Oscillations as a Probe of Quiescent and Erupting Solar Prominences" led by M. Luna.
\end{acknowledgements} 
 
%
   \bibliographystyle{aa} 
   \bibliography{promoscil} 
%
\end{document}